\documentclass[twocolumn,pra]{revtex4-1}
\usepackage[T1]{fontenc}
\usepackage[utf8]{inputenc}
\usepackage{amsmath}
\usepackage{amssymb}
\usepackage{graphicx}

\makeatletter
\usepackage{physics}
\usepackage{siunitx}
\usepackage{epstopdf}

\makeatother

\begin{document}

\title{Field ionization rate for PIC codes }

\author{I.\,Yu.~Kostyukov}
\email{kost@appl.sci-nnov.ru}
\author{A.\,A.~Golovanov}

%\selectlanguage{english}%

\affiliation{Institute of Applied Physics, Russian Academy of Science, 46 Uljanov
str., 603950 Nizhny Novgorod, Russia}
\begin{abstract}
An improved formula is proposed for field ionization rate covering
tunnel and barrier suppression regime. In contrast to the previous
formula obtained recently in {[}I. Yu. Kostyukov and A. A. Golovanov,
Phys. Rev. A 98, 043407 (2018){]}, it more accurately describes the
transitional regime (between the tunnel regime and the barrier suppression regime).
In the proposed approximation, the rate is mainly governed by two parameters: by the atom ionization potentials and by the external electric field, which makes it perfectly suitable for particle-in-cell (PIC) codes dedicated to modeling of intense laser--matter interactions. 
\end{abstract}

\maketitle
\begin{figure*}
	\includegraphics[width=1\linewidth]{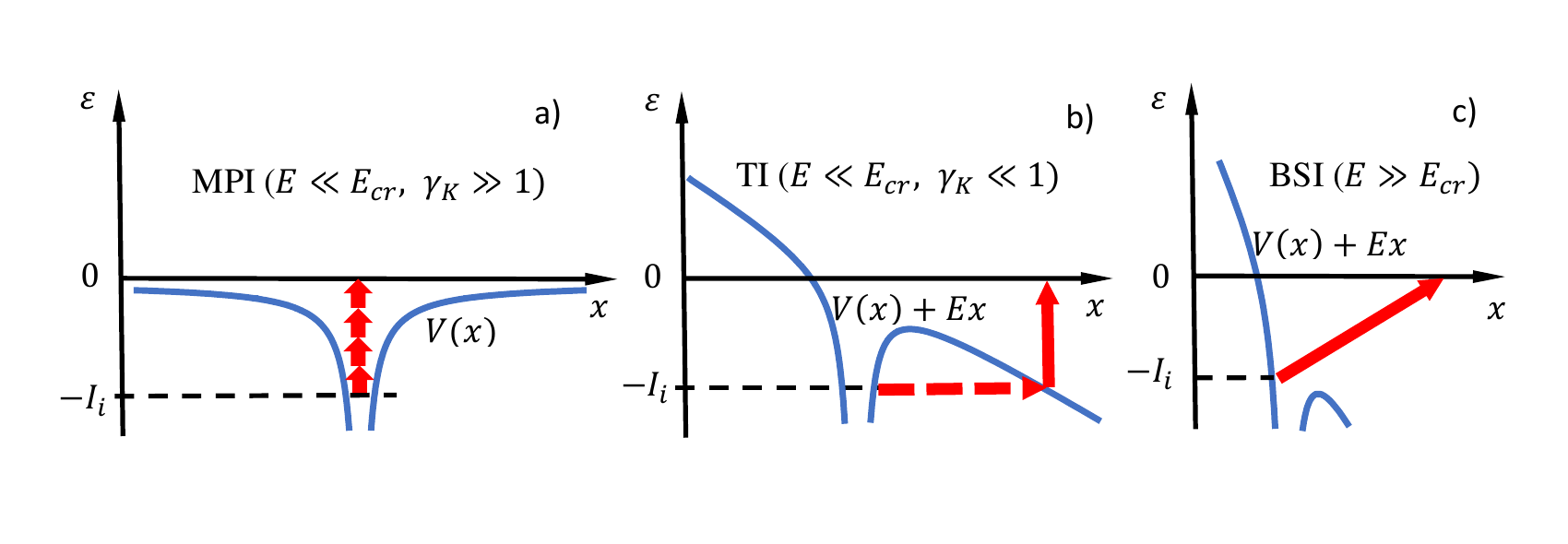} \caption{\label{regimes} Schematically, MPI (a), TI (b) and BSI (c) regimes of strong-field ionization in relation to the external field strength. Blue line demonstrates the atomic potential $V(x)$ or superposition of the atomic potential and the external field potential $V(x) + E x $. The red arrows show the electron transition during ionization. }
\end{figure*}

Ionization is one of the key processes in high-intensity laser--matter
interaction. The ionization-induced mechanisms play an important role
in many phenomena and applications like high-order harmonic generation
\cite{corcum,ivanov}, THz generation \cite{thz1,thz2,thz3}, ionization-induced
self-injection in laser--plasma accelerators \cite{Pak2010,McGuffey2010,Clayton2010},
triggering of QED cascades by seed electrons produced in ionization
of high-$Z$ atoms \cite{Tamburini2017,Artemenko2017}, etc. The ionization
in laser plasma can be caused by the collision of the atoms with the
energetic particles (impact ionization) or by action of the strong
electromagnetic field on the atoms (field ionization). The field ionization
includes roughly three regimes in relation to the electromagnetic field strength:
the multiphoton ionization (MPI) regime $E\ll E_{K}$, the tunnel ionization (TI)
regime $E_{K}\ll E\ll E_{cr}$ and the barrier suppression ionization (BSI)
regime $E\gg E_{cr}$ (see Fig.~\ref{regimes}), where $E_{K}=\omega(2m_{e}I_{i})^{1/2}/e$
is the field threshold associated with Keldysh parameter $\gamma_{K}=\omega_{L}(2m_{e}I_{i})^{1/2}/eE=E_{K}/E$,
$E_{cr}$ is the critical field above which the barrier of the atomic
potential is suppressed (it is defined quantitatively below). $I_{i}$
is the ionization potential of the atom (ion), $\omega$ is the laser
frequency, $c$ is the speed of light, $m_e$ is the electron mass, and $e > 0$ is the elementary charge.

The ionization rate can be calculated analytically in the multiphoton
and tunnel regimes \cite{Keldysh,Popov2004,krainov1998,USP2015}.
The tunnel ionization related to the strong field $\gamma_{K}\ll$1
can be treated in the static field approximation because the ionization
time (or the time of tunneling) is much shorter than the laser period
$\tau_{i}\simeq\gamma_{K}/\omega_{L}\ll1/\omega_{L}$ . The TI rate
in the static field approximation is \cite{USP2015,Perelomov1966-1,Ammosov1986}

\begin{eqnarray}
W_{TI}\left[F\left(E\right)\right] & = & \omega_{a}\cdot\kappa^{2}\cdot C_{kl}^{2}(2l+1)\left(\frac{2}{F}\right)^{2n^{*}}\left(\frac{F}{2}\right)^{m+1}\nonumber \\
 & \times & \frac{(l+m)!}{2^{m}m!(l-m)!}\cdot\exp\left(-\frac{2}{3F}\right),\label{Wpp}\\
C_{kl}^{2} & = & \frac{2^{2n^{*}}}{n^{*}\Gamma(n^{*}+l^{*}+1)\Gamma(n^{*}-l^{*})},\nonumber 
\end{eqnarray}
where $F=E/\left(\kappa^{3}E_{a}\right)$ is the normalized electric
field, $\kappa^{2}=I_{i}/I_{H}$, $n^{*}=Z/k$ is the effective principal
quantum number of the ion, $Z$ is the ion charge number, $l^{*}=n^{*}-1$
is the is the effective angular momentum, $l$ and $m$ are the orbital
and magnetic quantum numbers, respectively, $I_{H}=m_{e}e^{4}/\left(2\hbar^{2}\right)\simeq13.6$\,eV
is the ionization potential of hydrogen, $E_{a}=m_{e}^{2}e^{5}\hbar^{-4}\approx5.1\cdot10^{9}$\,V/cm
is the atomic electric field $\omega_{a}=m_{e}e^{4}\hbar^{-3}\simeq4.1\cdot10^{16}\,\mbox{s}^{-1}$
is the atomic frequency, $\Gamma(x)$ is the Gamma function \cite{Abramowitz}. 

When the external field is so strong that the maximum of the potential
barrier resulted from the superposition of the atomic field and the
external field is lower than the initial energy level of the electron,
the field ionization develops in the barrier suppression regime so that the electron becomes unbound and propagates above the barrier
instead of tunneling. In the BSI regime, the external field strength significantly
exceeds $E_{cr}=E_{a}\kappa^{4}/\left(16Z\right)$. It follows from
the estimations~\cite{kostyukov} that atomic electrons can be ionized in sub-PW laser pulses when  $E \gtrsim E_{cr}$ and the formulas for MPI and TI regimes are no longer applicable. Many empirical formulas for field ionization rates at $E \gtrsim E_{cr}$ were proposed \cite{Posthumus2018,Tong2005,Krainov1997,Bauer1999,Zhang2014}. Yet, most of them do not provide correct asymptotic in the high-field limit corresponding to the BSI regime. Moreover, they are applicable only for limited types of atoms and ions. 

Field ionization models are incorporated in many particle-in-cell (PIC) codes which
have become powerful and almost indispensable tools for the exploration of laser--matter interaction. Some models also include the energy losses associated
with ionization \cite{Rae1992,Nuter2011} and can be used to simulate
multiple ionization events within the time step of the PIC code main loop \cite{Artemenko2017,Nuter2011,Chen2013,Korzhimanov2013}.
Ideally, the formula for PIC codes should be simple and computationally
cheap, valid in a wide range of laser intensities, and applicable to all types of
atoms as well as all ion charges. Until recently, the field ionization
models used in the codes described only the TI regime or were based on too simple and inaccurate approaches.
For example, one of the model is based on the TI formula for $E<E_{cr}$ and the electron is assumed
unbound if $E\geq E_{cr}$ (see, for example, \cite{tamburini}). This model may dramatically overestimate
the ionization efficiency in BSI regime when the laser field is strong. 

Recently, the high-field limit of the BSI rate was calculated in the
classical \cite{Artemenko2017} and in the quantum \cite{kostyukov} approaches,
\begin{equation}
w_{BSI}(E)\approx0.8\omega_{a}\frac{E}{E_{a}}\sqrt{\frac{I_{H}}{I_{i}}}.\label{bsi}
\end{equation}
In this limit, the rate depends linearly on the external field strength
while the atomic system is characterized by the ionization potential
of the atom or ion. The piecewise formula for the field ionization
rate in both TI and BSI regime with correct asymptotic in the high-field
limit was also proposed \cite{kostyukov} 
\begin{equation}
w_{i}(E)\approx
\begin{cases}
W_{TI}(E), & E\leq E_{0},\\
W_{BSI}(E), & E>E_{0},
\end{cases}
\label{ti-bsi}
\end{equation}
where the field strength value, $E_{0}$, is determined from equation
$W_{TI}(E_{0})=W_{BSI}(E_{0})$. 

However, the accuracy of Eq.~\eqref{ti-bsi} is not high in the transitional
regime which corresponds to the laser field strength $E\sim E_{cr}$
and separates the TI and the BSI regimes. Here we suggest an improved rate
formula including besides the TI rate for $E\ll E_{cr}$ and the BSI
rate for $E\gg E_{cr}$ also the rate in the transitional regime $E\sim E_{cr}$.
The ionization rate near the atomic critical field $E\sim E_{cr}$
can be approximated by the empiric formula proposed by Bauer and Mulser for the hydrogen atom
\cite{Bauer1999}
\begin{equation}
w_{BM}(E)\approx 2.4 \omega_{a}\left(\frac{E}{E_{a}}\right)^{2}\left(\frac{I_{H}}{I_{i}}\right)^2.\label{bm}
\end{equation}
In contrast to $w_{BSI}$ it depends quadratically on the laser field
strength. The quadratic dependence of the ionization rate on $E$ and
transition to the linear dependence can be seen from Fig.~6 in Ref.~\cite{Bauer1999}
where the results of numerical integration of time-dependent Schrödinger
equation are presented.
Strictly speaking, the formula by Bauer and Mulser is applicable only to hydrogen and hydrogen-like ions with one electron and the charge of the atomic core equal to $Z$.
However, the formula is generalized to an arbitrary atom or ion by replacing $Z^2/2$ with $I_i$, and we expect it to provide a reasonable estimate for the ionization rate even in this case.
The resulting formula for the ionization rate
including the TI, the BSI, and the transitional regimes can be written as follows

\begin{equation}
w_{i}(E)\approx
\begin{cases}
    W_{TI}(E), & E\leq E_{1},\\
    W_{BM}(E), & E_{1}<E\leq E_{2},\\
    W_{BSI}(E), & E>E_{2},
\end{cases}
\label{ti-bm-bsi}
\end{equation}
where $E_{1}$ and $E_{2}$ are determined from equations $W_{TI}(E_{1})=W_{BM}(E_{1})$
and $W_{BM}(E_{2})=W_{BSI}(E_{2})$, respectively. The proposed formula
is well suited for PIC codes. It depends on the local instantaneous value of the
ionizing field strength as well as on the ionization potentials. 

\begin{figure*}[t!]
    \centering
	\includegraphics[]{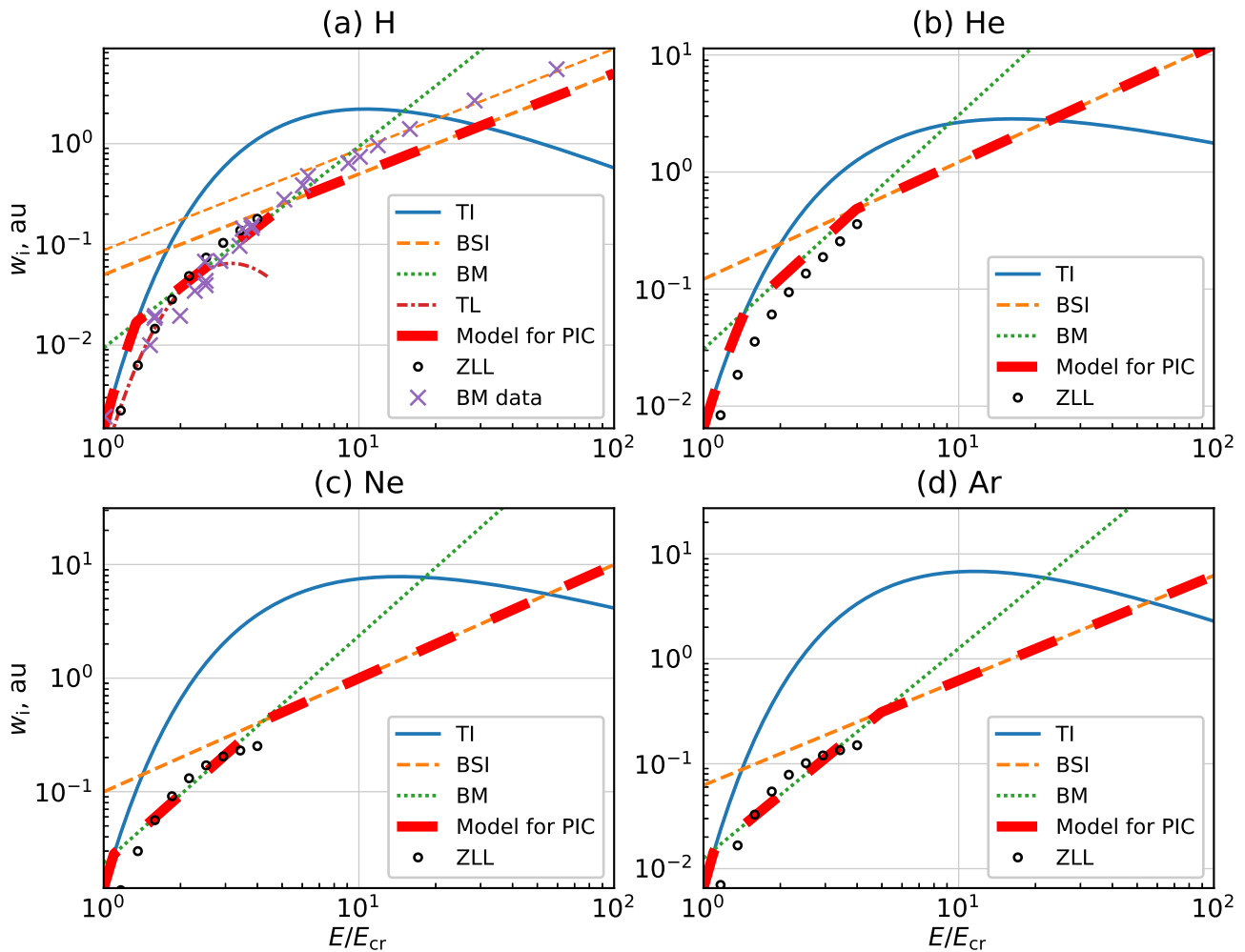}
    \caption{
        The field ionization rate for (a) hydrogen, (b) helium, (c) neon, (d) argon calculated numerically in Ref.~\cite{Bauer1999} (BM data, violet crosses) and Ref.~\cite{Zhang2014} (ZLL, black circles); calculated according to Eq.~\eqref{ti-bm-bsi} (red wide 
		dashed line); calculated according to Eq.~\eqref{bsi} for the BSI regime
		(orange dashed line); calculated according to Eq.~\eqref{ti-bsi}
		for TI regime (blue solid line); calculated according to Eq.~\eqref{Wpp}
		and proposed in Ref.~\cite{Bauer1999} (BM, green dotted line); proposed
    	in Ref.~\cite{Tong2005} (TL, red dashed-dotted line).
        The narrow dashed line in (a) corresponds to Eq.~\eqref{bsi} but with numeric coefficient of $1.4$ instead of $0.8$.}
    \label{atoms}
\end{figure*}

First we compare the predictions of the proposed formula~\eqref{ti-bm-bsi}
for hydrogen with the numerical results obtained in Ref.~\cite{Bauer1999}
by solving the time-dependent Schrödinger equation (see Fig.~\ref{atoms}a).
It is seen from Fig.~\ref{atoms}a that the analytical and the numerical
results are in a fairly good agreement. The dependence in the BSI regime is indeed linear in the numerical simulations, but the numerical coefficient is different resulting in a small displacement when plotted in the logarithmic scale.
This difference may be attributed to imprecise definition of the ionization rate.
Unlike the TI regime, the dependence of the total ionization probability on time in the BSI regime is not exponential (See. ref~\cite{kostyukov}), and therefore the instantaneous probability does not depend on the instantaneous field value but on its history.
Introducing $w_i(E)$ is thus an approximation used to qualitatively describe the ionization rate.
Some of the ways of determining the numeric coefficient in this dependence which lead to slightly different results are discussed in Ref.~\cite{kostyukov}.

Similar comparisons
were done for helium, neon and argon atoms (see Fig.~\ref{atoms}b--d).
The numerical data is obtained by integration of the Schrödinger equation
in the single-active-electron approximation and provided in Ref.~\cite{Zhang2014}.
The analytical and numerical results are also in a good agreement. 

In conclusions, we have proposed the improved formula for strong-field ionization
rate covering a wide range of laser intensities from the TI regime to the BSI regime. The formula is well suited for PIC codes as it depends on the local instantaneous value of the ionizing field strength while the dependence on the atomic systems
is expressed via the ionization potentials. Therefore, it is applicable for all types of atoms as well as for all ion charges, and it is computationally cheap. The formula predictions are in good agreement with the results of numerical simulations of field ionization. 

The research was supported in part by the Ministry of Science and Higher Education of the Russian Federation (state assignment for the Institute of Applied Physics RAS, project No. 0035-2019-0012) and by the Foundation for the Advancement of Theoretical Physics and Mathematics “BASIS” through Grant No. 17-11-101.

\end{document}